\documentclass[preprint,superscriptaddress,aps,pra]{revtex4-1}

\usepackage{color}  

\usepackage{graphicx}

\begin{document}


\title{Why free markets die: An evolutionary perspective}


\author{E. Viegas}
\affiliation{Complexity \& Networks Group and Department of Mathematics, Imperial College London, SW7 2AZ, United Kingdom}

\author{S. P. Cockburn}
\noaffiliation

\author{H. J. Jensen}
\affiliation{Complexity \& Networks Group and Department of Mathematics, Imperial College London, SW7 2AZ, United Kingdom}

\author{G. B. West}
\affiliation{Santa Fe Institute, 1399 Hyde Park Road, Santa Fe, NM 87501, United States}


\date{\today}

\begin{abstract}

Company mergers and acquisitions are often perceived to act as catalysts for corporate growth in free markets systems: it is conventional wisdom that those activities lead to better and more efficient markets. 
However, the broad adoption of this perception into corporate strategy is prone to result in a less diverse and more unstable environment, dominated by either very large or very small niche entities.
We show here that ancestry, i.e. the cumulative history of mergers, is the key characteristic that encapsulates the diverse range of drivers behind mergers and acquisitions, across a range of industries and geographies.
%
A long-term growth analysis reveals that entities which have been party to fewer mergers tend to grow faster than more highly acquisitive businesses.

\end{abstract}

\pacs{}

\maketitle


Mergers and acquisitions of firms and companies are fundamental activities that dynamically shape the ecosystems of the business world.
%
Conventional financial studies tend to use financial and economic data in order to explain these dynamics,  
however the benefits of this activity remain uncertain, despite a significant amount of research \cite{Kwan1999,Madura1994,Toyne1998}.
%
In this paper, our approach instead combines influences from two areas which have recently been successfully drawn upon to analyse the structure and stability of financial systems, the complementary fields of complex networks \cite{Ueno2007} and evolutionary dynamics \cite{Viegas2013}, with particular reference to the relationship between diversity reduction
and increasing systemic risk \cite{Haldane2011}.
Specifically, we translate to a business context the important distinction between phenotype and genotype: the former is associated to externally observable attributes, which are influenced by a combination of genetic, environmental and  developmental factors; the latter instead refers to the more fundamental, underlying genetic instructions which hold the design for these attributes \cite{Jablonka2005}.
%
%
We argue that the reason for the lack of consensus between existing, traditional financial analyses is simple: financial data gives a snapshot of the condition of the markets, i.e it is a measure of the emergent market state and therefore analogous to the phenotype. It is not however the underlying driver behind the dynamics shaping the markets at a fundamental level. 
Our findings show that it is in fact \emph{ancestry} - the cumulative number of mergers of all acquired entities - that is the key quantity for determining the probability for an institution to be the originator of a merger, and in this sense is analogous to the genotype, which encodes the past evolutionary process. 	

Although a range of factors might be imagined to feature in the merger and acquisition process - such as the regulatory environment, balance sheet size of a business, potential synergies across different agents, value of a brand etc. - surprisingly, it is ancestry that captures the essence of this dynamics. Moreover, we find this to be quite general across markets from several countries and industries, indicating the fundamental role that ancestry plays, and that the ancestry-based mechanism may have a wide, universal application across capitalist systems.
To analyse the observed importance of ancestry, we devise a simple agent based model for the mergers and acquisition of companies, which is indeed found to accurately describe data taken from diverse markets.

A consequence of an ancestry-based mechanism underpinning the mergers and acquisitions process, is a size-wise bi-modality in the types of entities that exist, which results in an imbalanced ecosystem, mostly consisting only of either a few very large - ``too big to fail'' - entities, or very small, niche entities. 
The emphasis on sustained growth from larger entities tends to result in the depletion of mid-sized entities, as the latter become targets of mergers, which ultimately leads to a vacuum in this section of the industry. The ancestry-based mechanism also inhibits the introduction of new mid-sized market entrants, as new entities are typically either incorporated into larger existing entities, or remain within the small, niche pool.
%
The establishment of entities that are ``too big to fail'' ultimately creates a paradigm for free market systems whereby the prospect of nationalization - directly or through subsidies - becomes a potentially feasible outcome.


\subsection{Ancestry data analysis}
\label{sec:dataAnalysis}
%

The balance sheet of a company is, of course, a key measurement of the state of a business, and hence is often considered to be an important tool for predicting future acquisitions. 
Indeed, a number of academic studies indicate 
that larger entities are more likely to be acquirers \cite{Becalli2013}. Given that there is strong correlation between balance sheet size and ancestry, akin to phenotype and genotype respectively, it is therefore not inconsistent with traditional finance theory when we find that ancestry plays a crucial role in predicting acquisitions. We are even able to clarify the relationship between ancestry and balance sheet: while balance sheet size is indeed a measurement of the state of a business, ancestry is the fundamental property that drives the dynamics of merger and acquisitions between companies, which then result in changes to balance sheet size. 

\begin{figure}[h!]
	\includegraphics[scale=0.42, clip]{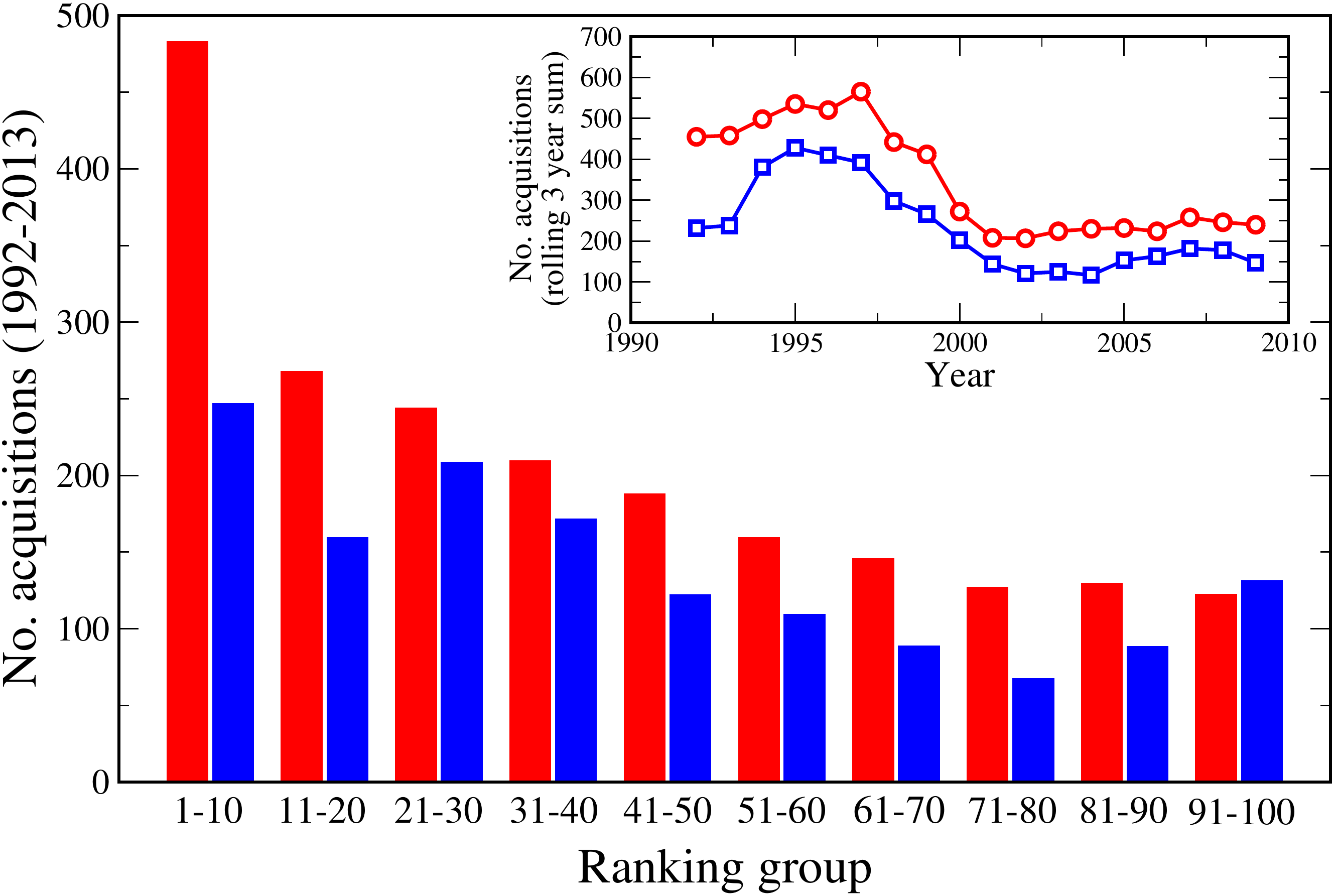}
	\caption{\it {\bf Comparison of ranking methods to predict merger activity (US banks): Ancestry vs. Balance Sheet.} Entities are ranked based on ancestry (red) or balance sheet size (blue); the bar chart shows the time averaged number of mergers between banks (vertical axis) by ranking group (horizontal axis); for each ranking group, the number of acquisitions is averaged over a three year period, starting at each year between $1992$ and $2013$. The ranking by ancestry leads to a monotonic decrease by ranking group, and a greater number of mergers, than ranking by balance sheet size. Inset: Number of mergers in a $3$ year window from year indicated on horizontal axis for top $100$ banks, ranked by ancestry (red) and balance sheet (blue). The top group ranked by ancestry consistently underwent more mergers than the top group ranked by balance sheet size.}
		\label{fig:BSvsAnc}
\end{figure}

Using data from the US banking industry (see Methods Summary), we demonstrate in Fig.~\ref{fig:BSvsAnc} that ancestry provides a consistently better metric than size (measured in this case by balance sheet) for predicting mergers and acquisitions for these banks.
The main plot in Fig.~\ref{fig:BSvsAnc} shows the number of acquisitions for US banks as time averages (i.e. gathering data over a time window and averaging the readings) over successive three year windows between $1992$ and $2013$, comparing two ranking methods: (i) ranking by number of ancestors and (ii) ranking by balance sheet size. 
To obtain this plot, all banks are first ranked by both metrics within each year; the average number of acquisitions in the following $3$ years is then calculated. These acquisition numbers are aggregated by ranking group, so the data point for each ranking group shows the sum of all merger activities within that group of banks, over all times. By comparison, the inset shows the total number of acquisitions as a function of time. Specifically, the data here shows the mergers that occurred across the top $100$ banks, within a $3$ year window from the year referenced on the x-axis that the data point sits at. 

From Fig.~\ref{fig:BSvsAnc}, we see that the data for the ancestry ranking shows a far more consistent behaviour than the balance sheet ranking, displaying an almost entirely monotonic decay with increasing rank; i.e. those banks ranked higher by ancestor number subsequently partook in more mergers, as would be expected of an effective ranking method. The balance sheet ranking instead shows a less regular behaviour and, moreover, captures far fewer merger events across bank ranking groups, as shown in the main figure, and over all times, as shown in the inset. This leads to the following conclusions: firstly, entities with many ancestors are generally more likely to be involved in merger activities; secondly, ranking by ancestry places the banks in a more accurate order, in terms of merger likelihood, than ranking by balance sheet size.

\begin{figure}[h!]
	\includegraphics[scale=0.45, clip]{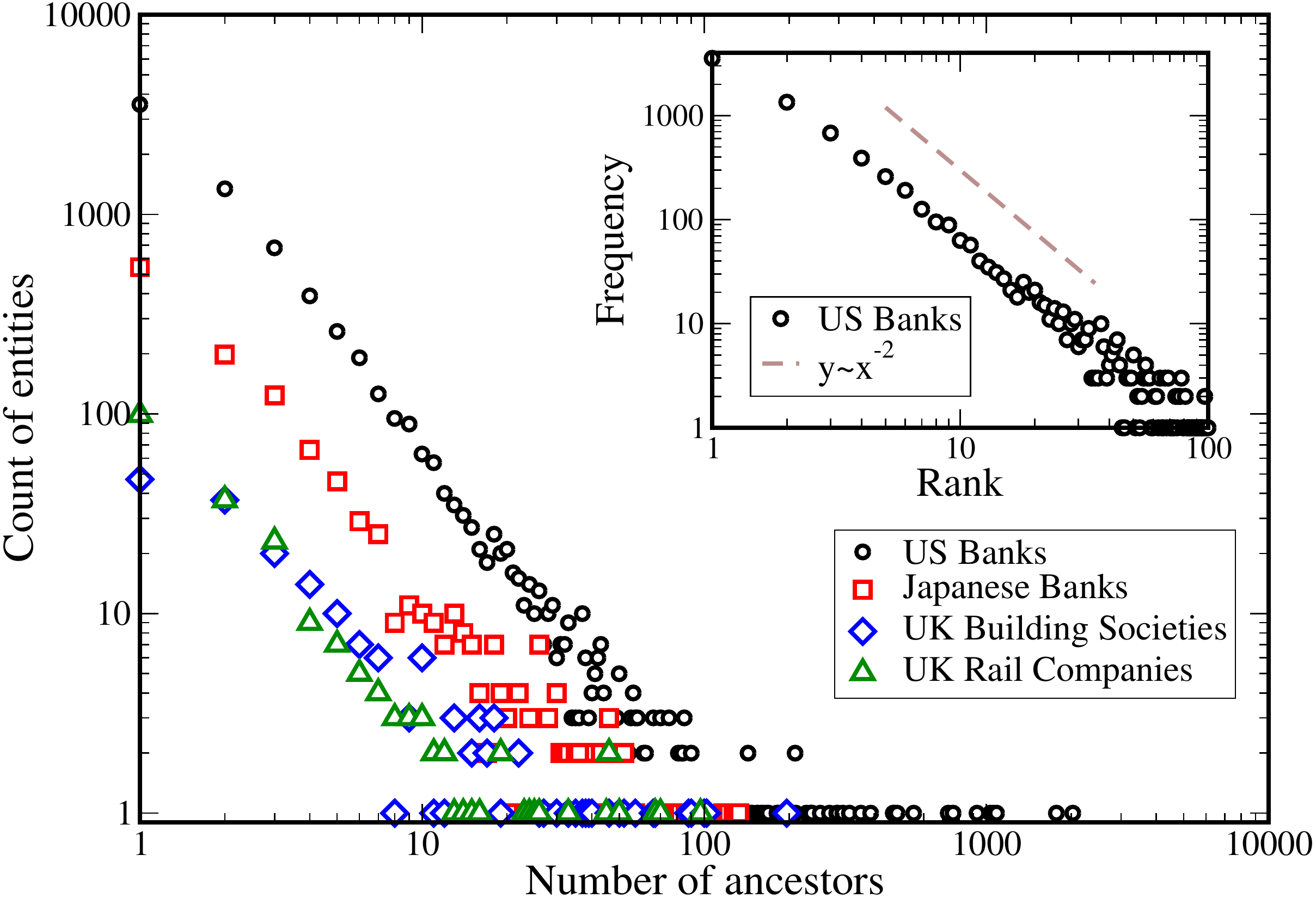}
	\caption{\it{\bf Structure of ancestry data.} Log-log plot showing distributions of ancestry numbers across several industries and countries. Inset: Zipf plot for US Banking ancestry data; the dashed line has a gradient of $-2$.}
	\label{fig:AllDataPlusZipf}
\end{figure}

Given the importance of ancestry to merger activities in the US banking sector, 
we have also expanded the study to examine different geographies, time periods and industries.
Specifically, we move on to explore the structure of business ancestries by additionally considering: US state-level banks, Japanese banks, UK building societies and UK rail companies, over various time periods. 
The main focus in each case is the distribution of businesses by ancestry size; i.e, the distribution of businesses ranked by their number of ancestors,
which is shown in Fig.~\ref{fig:AllDataPlusZipf}. There is a striking similarity between all the observed distributions:  they all exhibit very similar power-law like behaviour. 


\subsection{Comparison between ancestry data and agent based model}
\label{sec:results}

\begin{figure}[h!]
	\includegraphics[scale=0.4, clip]{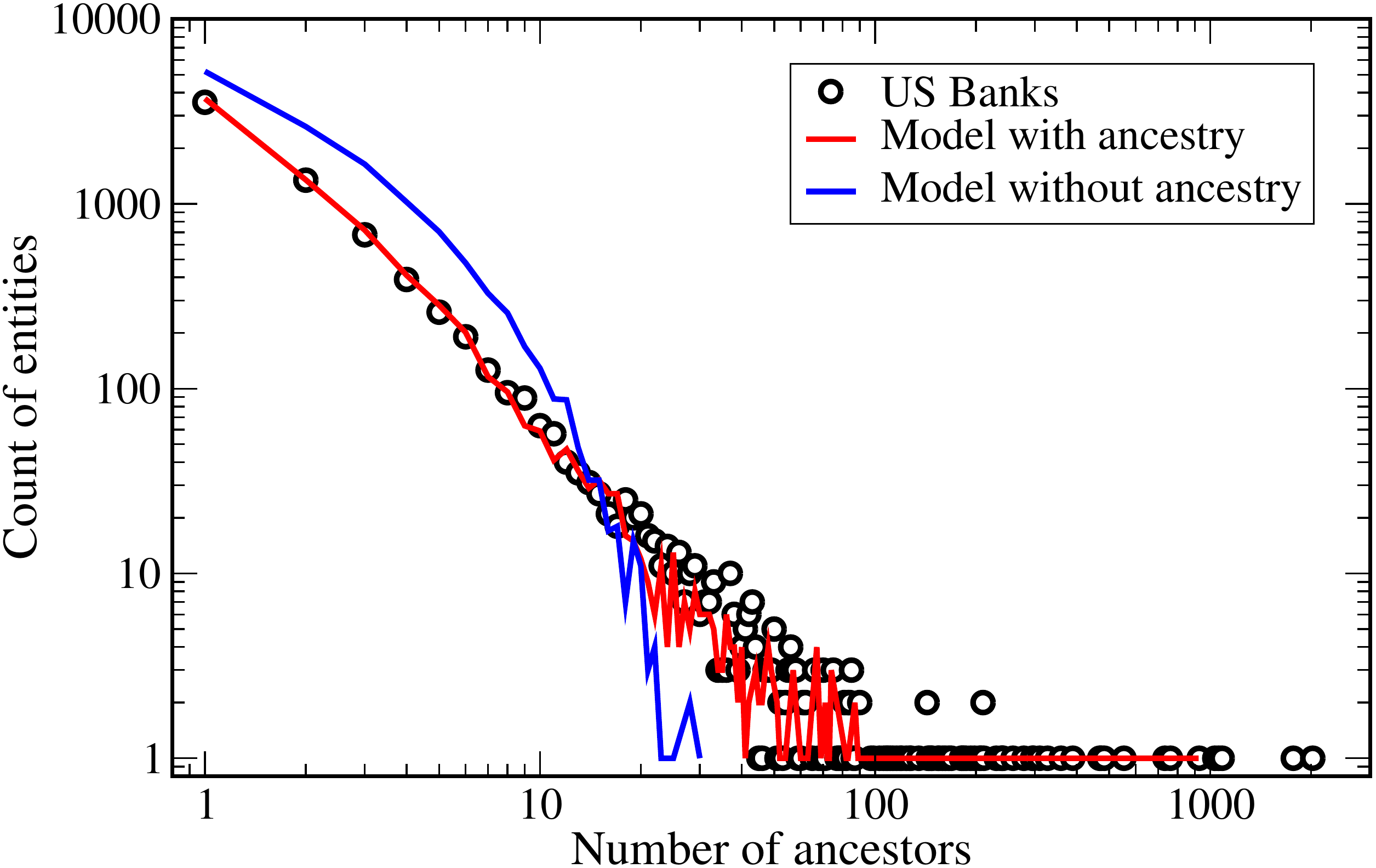} 
	\caption{\it {\bf Importance of ancestry weighting to dynamics.} A comparison is shown between US banking ancestry data and model simulations with (blue) and without (red) an ancestry weighted merger probability. Clearly the former leads to a far closer agreement with the US data.}
	\label{fig:randomWalk}
\end{figure}
We  develop a simple agent based model for a market in which mergers occur, and compare this initially to the US banking data set (see Methods Summary for details of data and model). Our model is based upon Price's Cumulative Advantage Theory \cite{Price1976} for the bibliometric process, and Simon's stochastic process formulation \cite{Simon1955}. The sole agents of the model are represented as a vector, ${\bf B}=[B_{0}, B_{1,\ldots}]$, which undergo cycles of dynamics.
Crucially, the probability of merger for each agent is weighted by their number of ancestors, $A_{\rm n}$, as follows:
\begin{equation}
	p_{\rm merger}=p(1+A_{\rm n})^{3/2},
	\label{eq:eq1}
\end{equation}
this specific functional form can be considered as a consequence of the Zipf plot \cite{Zipf1942} of Fig.\ref{fig:AllDataPlusZipf}.

The first key point to highlight is the non-stationary, evolving nature of the mergers and acquisitions process for companies,
which is clearly demonstrated in Fig.\ref{fig:randomWalk}; here we compare the results of a single realisation of the full model (red line) to the actual accumulated ancestry data for the US nationally (black circles), and contrast those with the results of a ``random walk'' model (blue line), where in the latter all agents have equal, and \emph{time independent}, probability to merge (this corresponds to setting $A_{\rm n}=0$ in Eq.\ref{eq:eq1}).
The ``random walk'' model leads to a noticeable deviation from the US data and a loss of the power-law structure of the ancestry distribution, while a merger probability, which depends on the past history through the ancestry size yields excellent agreement across the entire range. We conclude that the size of the ancestry can be seen as encoding how keen a company is to originate a merger, i.e. the ancestry  record encodes a phenotypical trait.

%

%

\begin{figure}[h!]
	\includegraphics[scale=0.4, clip]{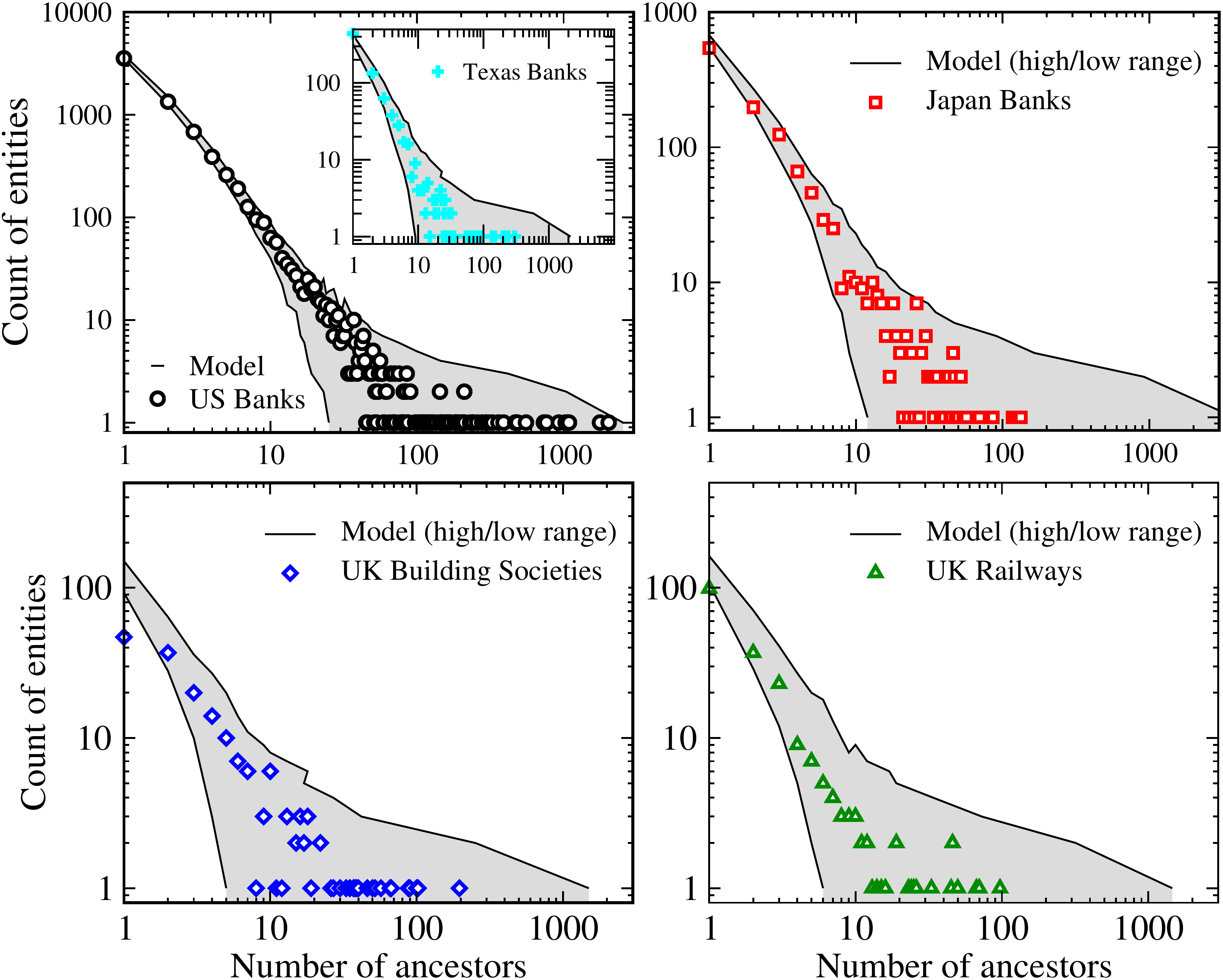} 
	\caption{\it{\bf Model vs. real world data.} Mergers data is shown for US banks ($1970$-$2013$), Japanese banks ($1872$-$2013$), UK building societies ($1936$-$2012$), and UK railway companies (late $1700$s-$1923$). In each case the model predictions are shown by the grey shaded region, which indicates the range of predictions across $1000$ simulations. The inset shows the subset of US data for banks within Texas.}
	\label{fig:USandJapan}
\end{figure}

To further test the model, Fig.~\ref{fig:USandJapan} shows a  detailed comparison to several additional sets of data: the US (national and state-specific; $1970$-$2013$) and Japanese ($1872$-$2013$) banking sectors, and UK building societies ($1936$-$2012$) and railway companies (late $1700{\rm s}$-$1923$). 
Given the very distinct markets, industries, and time-lines covered by these data sets, the close agreement between model and results suggest that these differences are of little consequence for the merger and acquisitions process over a long term evolutionary period, and, importantly, that an ancestry-based model captures the fundamental and essential dynamics across these systems.
It is interesting to note that the regulatory framework that exists in the US is fragmented in nature, as different banks need to comply with regulations that can be federal banking statutes and/or state specific laws. We can therefore consider the similarity between state and federal level as indicating a certain robustness to the ancestry mechanism. And indeed, the distinctive legislation, as well as time period of acquisitions, does not seem to bear any impact on the long term evolutionary process.

Restricting the analysis to individual US states, we also find a favourable agreement between model and data set.
This suggests that there exists a kind of self-similarity between subsets of the systems (e.g. the individual states, in the US case) and the entire system (all US banks);
an example from the largest banking state, Texas, is shown in Fig.~\ref{fig:USandJapan} (upper left inset). 
Of particular relevance to the state wise analysis, is the Riegle-Neal Interstate Banking and Branching Efficiency Act, which came into force in $1994$; 
previously, the $1982$ Garn-St. Germain Depository Institutions Act had allowed interstate mergers and acquisition only for failed banks \cite{Sherman2009}.
It is therefore fascinating that both the inter- and intra-state merger data shown in Fig.~\ref{fig:USandJapan} demonstrate a very similar behaviour.

\subsection{Growth through mergers}
\label{sec:growth}
\begin{figure}[h!]
	\includegraphics[scale=0.35, clip]{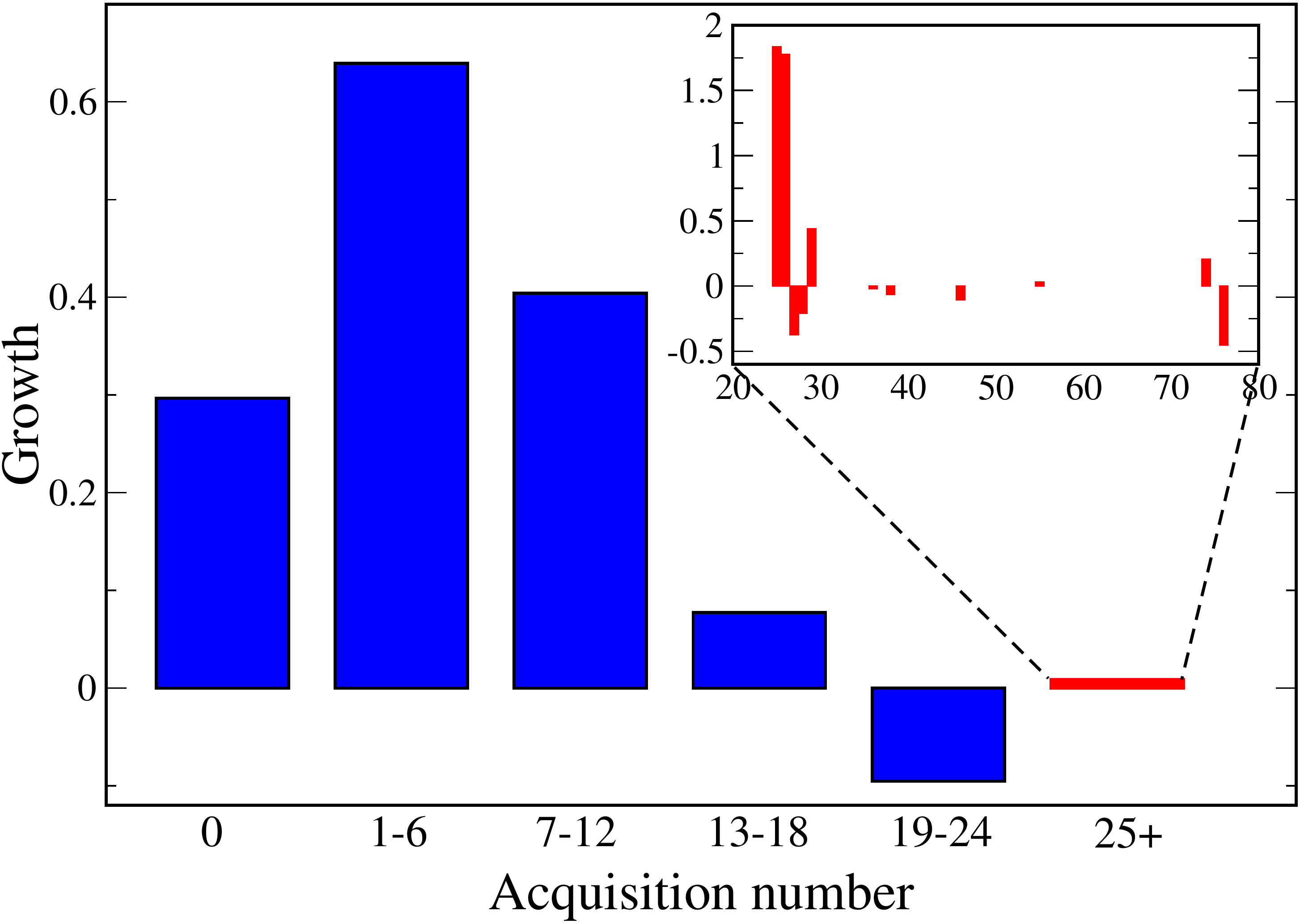}
	\caption{\it{\bf Real business organic growth.} The chart shows the actual, organic growth of entities within each acquisition grouping, excluding the effects of GDP growth and balance sheet aggregation from mergers between US banks from $1992$ to $2013$. For each of the surviving banks in $2013$, we aggregated the balance sheets of each of their ancestors in $1992$, and indexed by the GDP. As a result, the real organic growth is derived from the ratio between the balance sheets of surviving entities as at $2013$, and those of their respective ancestors (GDP linked and aggregated) as at $1992$.
	The scale is relative to GDP such that values $>0$, $=0$, and $<0$ indicate growth greater than, tracking, and less than GDP, respectively.
	The red bar in the main plot shows the average over those banks with ancestries $>25$; the inset shows a refined view of growth for these largest banks, by number of acquisitions. Note that in combining the inset data in the main plot, the average is weighted by balance sheet size.}
	\label{fig:BSBarChart}
\end{figure}
Given that merger dynamics can be described simply in terms of ancestry, financial data features instead as a result of these processes, rather than as inputs driving acquisitions. There arises a question then, as to how financially beneficial such processes are to individual businesses. To analyse this, Fig.~\ref{fig:BSBarChart} shows an analysis of the growth of businesses, as a function of acquisition number.

It is clear from this figure that growth levels do not increase, on average, for entities having undergone numerous mergers, and they instead remain GDP trackers. This indicates that on average the largest businesses do not realise benefits from acquisitions, however individual cases can lead to a marked growth.

\subsection{Dominance of Largest entities}
\label{sec:mergers}

To explore the effect of the ancestry-based merger model upon a market as a whole, we consider now the changes in total assets across different bank sizes, ranked by centiles. 
%
\begin{figure}[hc!]
			\includegraphics[scale=0.48, clip]{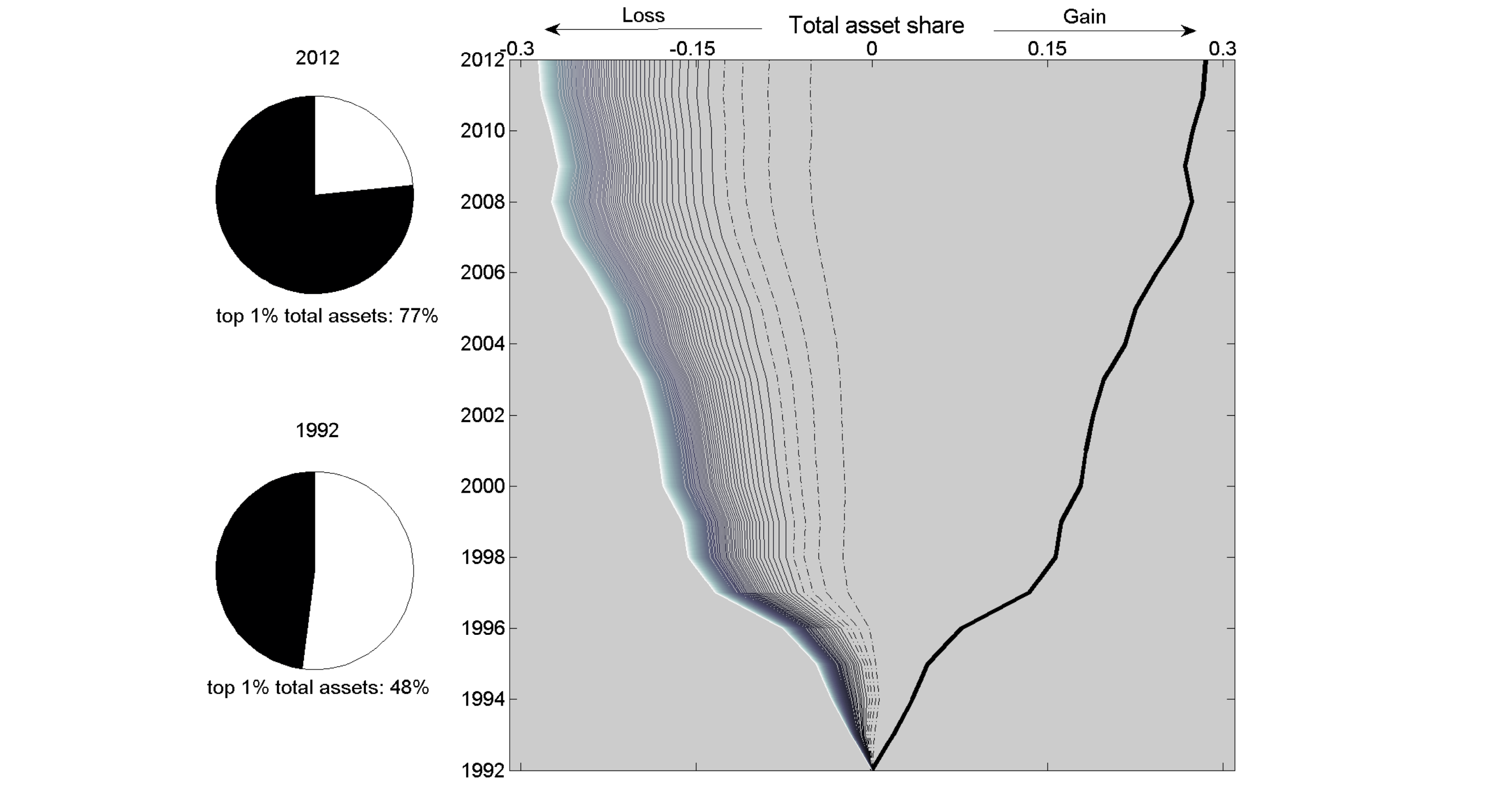}
			\caption{\it {\bf Size-wise bi-modality in US banks.} The plot shows the change in the market share of US banks over time, measured by growth in fractional balance sheet size.
			The banks are ranked into percentiles by proportion of total balance sheet size, with the top $5\%$ isolated for clarity in the plot: the rightmost, thick black line is the top $1\%$; the next $4$ percentiles are shown by dot-dashed lines. Moving to the left, the remaining lines represent each of the the remaining percentiles.
			The $2^{\rm nd}-100^{\rm th}$ percentile lines are plotted in a cumulative sense, to illustrate that the total loss from all of these market sections is equal to the growth in the largest $1\%$ of the market. The gap between each of the $2^{\rm nd}-100^{\rm th}$ percentile lines, therefore indicates the size of the loss for each percentile. The plot shows that the gap, and so loss, is largest for the $2^{\rm nd}\sim 10^{\rm th}$ percentile range, and decreases for subsequent percentiles, illustrating the mid-sized part of the market has suffered the largest loss by $2012$.
			}
		\label{fig:bifurcation}
\end{figure}
Focussing upon the US banking sector, Fig.\ref{fig:bifurcation} shows the change in market share based on balance sheet data from $1992$ to $2012$.
The figure shows a clear increase in the total assets of the largest $1\%$ of banks, partnered with a reduction in the total assets held by entities from all other percentiles.
The increase in total assets is almost exclusively concentrated within the largest $1\%$ of banks (thick black line), while the corresponding reduction in total assets comes mainly from the percentiles $2-10$, with the remaining higher percentiles (and so smaller entities), contributing far less. This shows that it is the total assets of the top $2\%-10\%$ banks by size, or mid-sized section of the market, that are depleted most obviously in order to fuel the growth of the largest banks.
%

It is the mid-sized section of the market that sees the greatest impact. This happens because the ancestry driven mechanism underlying bank mergers and acquisitions affects this  part of the market in two ways: there are fewer agents in this size range the smallest agents (due to their abundance) and the mid-size agents are less aggressive acquirers than the largest entities. Ultimately, this situation effectively leads to a bimodal population, in which only either very large entities or very small entities survive.


%
%

\subsection{Discussion and Conclusions}

The dynamics of mergers and acquisitions between businesses are, at a fundamental level, more strongly linked to ancestry than balance sheet size. 
A simple agent-based model, incorporating the dependence of merger probability upon ancestry, was found to give excellent agreement with merger and acquisitions data obtained from a range of time periods, industries and geographies.
While it is remarkable that the structure of the real world data can be reproduced from such straightforward principles, this indeed suggests that a  merger probability which increases with ancestry is a crucial part of the mechanisms behind the observed data. 
We have found that long-term growth is not, on average, markedly better than GDP growth for businesses that have undergone many mergers.
Moreover, the ancestry driven nature of mergers is found to lead to a reduction in market diversity and an exposure to the associated systemic risks \cite{Haldane2011}. 

The findings above provide three key insights of broader relevance to free markets systems:
Firstly, traditional financial analysis would benefit from utilising non-financial as well as financial data in assessing market dynamics in the same manner that biological systems benefit from harmonious analysis of interactions between genotype and phenotype.
Secondly, from a regulatory perspective, our findings suggest that it is key that legislation designed to bolster stability is not solely focused on the acquisitions among large entities, but should also protect the organic growth of smaller, rapidly expanding enterprises. 
Finally, while counter-intuitive, the findings that mergers do not contribute to an enhancement of growth for entities with very large ancestries naturally opens a debate as to whether, as in the world of biology, companies' growth follows a sigmoid curve, meaning there is a natural size limit beyond which growth is no longer advantageous.

\section{Methods summary}
\label{sec:theory}

\subsection{Agent based model}
\label{sec:model}

The basic aim of the model to capture the ancestry (i.e. genotype) driven merger dynamics that underlie observable financial data, the phenotype of this analogy.
%

%
%
The core probability of Eq.~\ref{eq:eq1}, $p$, is fixed throughout all cycles at $p = 1/40000$, however a large range of values yield very similar results.
Note also that setting $A_{n}$ to zero would yield a purely random model, i.e. the random walk model of fig~\ref{fig:randomWalk}.
In each cycle, Source banks are selected based on the ancestry-based probability of Eq.~\ref{eq:eq1}, after which each randomly selects a merger partner; these partners cease to exist in the live population and the Source Bank adds the number of ancestors $A_{n}$ to its number of ancestors.
The initial number of agents is obtained directly from the data set being modelled and each agent is deemed to have no ancestors on model initialization.
Cycles continue until the total number of remaining agents is equal or less than the corresponding number in the final time input data.
The model data shown are based on $1000$ repeated simulations.

Let us remark on how the dependence on ancestry relates to the Preferential Attachment concept\cite{Simons1955,Barabasi1999}  
and  Cumulative Advantage\cite{Newman2010}.
In relation to our study these mechanisms are clearly distinct from each other.
The term Preferential Attachment is used to describe a process where the new, and smaller, vertices express a preference towards attaching to larger vertices. Therefore, the behavior of the agent - the new arriving vertex - is determined by  features of another agent. In contrast, Cumulative Advantage is a process where an inherent characteristic of an agent drives the behavior of that same agent. The dependence of the merger probability on ancestry is in this sense an example of ``Cumulative Advantage'' and distinctly  different from the Preferential Attachment  mechanism.



\subsection{Data sources}
\label{sec:dataSources}

The data sources used in this work were: (i) 
the Federal Deposit Insurance Corporation (FDIC) Institutional Directory, (ii) the Japanese Banking Association (JBA) data directory, (iii)
the UK Building Societies Association (BSA) member files and (iv) data on UK rails companies contained in Awdry's {\it Encyclopaedia of British Railway Companies} \cite{Awdry1990}.

\subsection{Acknowledgement}

HJJ is supported by  the European project CONGAS (Grant FP7-ICT-2011-8-317672).

\end{document}